\documentclass[a4]{article}

\usepackage{makeidx}     
\usepackage{graphicx}    
\usepackage{multicol}    
\usepackage{amsmath}

\begin{document}
\title{Market Simulation Displaying Multifractality}

\author{Kazuko Yamasaki and Kenneth J. Mackin\\
Tokyo University of Information Sciences,1200-2,Yatou
town,Chiba,Japan\\
yamasaki@rsch.tuis.ac.jp}
\maketitle
\begin{abstract}
We proposed a market simulation model (micro model) which displays multifractality and reproduces many important stylized facts of speculative markets. From this model we analytically extracted the MMAR model (Multifractal Model of Asset Returns)\cite{Mandelbrot0} for the macroscopic limit.
\end{abstract}
\section{Introduction}
\textbf{Current Studies by Market Simulations}\\
Many studies have tried to reproduce the stylized facts of speculative markets by means of market simulation with the aim of understanding the true nature of speculative markets. Stylized facts studied include (1)fat-tail, (2)long-memory of nonlinear function of returns, (3)short-memory of raw returns and (4)the time scale invariance property. In addition, recently many empirical investigations have shown the multifractality of time series of speculative markets \cite{Ausloos1}. These stylized facts are very characteristic and universal so that we expected if the stylized facts could be reproduced by a market simulation, we could narrow the cause of these facts in order to analyze the cause and effect. Recently many market simulations have succeeded in reproducing some of the facts, but these reports have given different or even contradictory explanations. The different results between reports comes from the nature of multi-agent simulations, which includes a lot of freedom. Though we believe the fractality (or multifractality) has been one of the most important feature of the speculative markets since first mentioned by Mandelbrot, most market simulations have not examine this aspect. This research is based on the following point of view.\\
\textbf{Extraction of Macro Model from Micro Model}\\
In this research, we define the model which includes individual variables for each agent ``the micro model''. (e.g. Each trader's parameters used in buy-sell decisions.) We define the model which is described only with macro variables such as price and volume ``the macro model''. We belive such extraction is important for understanding of the mechanism of speculative market, as it was first emphasized by Takayasu et. al.\cite{Takayasu1}. This also makes the correspondence between the result and the cause clear. The following table compares 3 previous macro model stochastic equations.
\begin{center}
\begin{tabular}{l|c|c|c}
\hline
\hline
 & GARCH & FBM($H\ne0.5$) & MMAR\\
\hline
(2)long memory &   no & yes & yes\\
(3)short memory & yes & no &  yes\\
(4)fractality  &   no & yes & yes\\
\hline
\hline
\end{tabular}
\end{center}
As seen in the table, The GARCH model and FBM (Fractional Brownian Motion) model can reproduce only a part of the stylized facts. The MMAR model \cite{Mandelbrot0}, proposed by B. Mandelbrot can reproduce most of the main stylized facts. In addition, the model can show both long-memory and Martingale property simultaneously.\\
\textbf{MMAR Model}\\
MMAR model can be described as follows.
\begin{gather}
\begin{split} 
\log{P(t)}-\log{P(0)}=B_H[\theta(t)]\\
\end{split}
\end{gather}
Here, $B_H(t)$ is a fractional Brownian motion with self-affin index $H$, and $\theta(t)$ is a stochastic trading time which is a cumulative distribution function of  multifractal measure. (The trading time is counted every time a trade occurs.)  \\
An example of multifractal measure is
\begin{gather}
\begin{split}
\mu= \prod_{k=0}^{K} M_k
\end{split}
\end{gather}
Here, $M_k=M(\eta_1,\eta_2,...\eta_k)$ is a stochastic variable which changes with the kth time scale. 
$B_H(t)$ and $\theta(t)$ are assumed to be independent.
The meaning of this model is that the price changes under fractional Brownian motion along with the trading time, but the trading time shows multifractality.  
\section{Proposed Market Model}
In this model, we aimed at analytically extracting the macro model from the micro model sacrificing some reality.\\
\textbf{Time Scale}\\
In our daily lives we commonly apply many layers of time scales. \\
\includegraphics[scale=0.3]{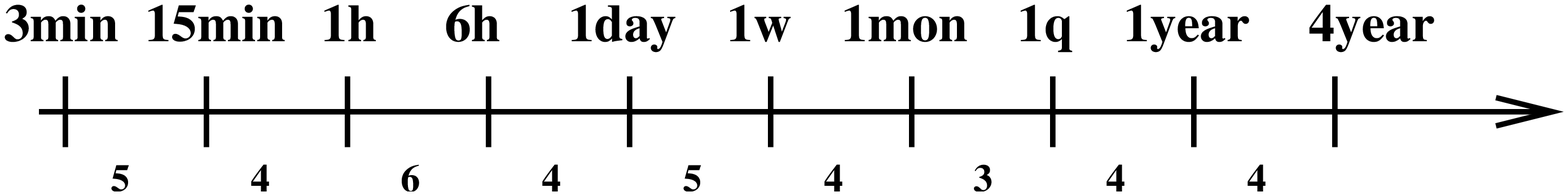}\\
We supposed that the role of many layers needed for multifractality was carried by the different information in each layer. Each trader decides his judgment based on such information. In this model, all the information comes from external sources (e.g. a macro index such as GDP). This is because price and trading time are independent in the MMAR model, and if a trader's judgment decides the trading time, he cannot refer to the prices for his judgment.\\
\textbf{Trader Judgment}\\
$M_k$ is the strength of information for source $k$. Generally if a trader's judgment consists of the combination of the logical products (and) and the logical sums (or) of several sources of information\\  
 (($M_1$ and $M_2$) or  $M_3$ ) and ($M_4$ or $M_5$ ) and $M_6$\\
this can be rewritten in the form\\
 ($M_1$ and $M_2$ and $M_4$ and $M_6$) or \\
($M_3$ and $M_4$ and $M_6$) or ... etc.\\
We supposed a trader makes an order in single shares with a probability in proportion to the strength of his reference sources of information. Here, we do not distinguish between buying order and selling order. 
The strength of the source of information $M_k$ is a stochastic variable, and $m_k$ is its sample value.
We define the probability of a trader making an order, as follows.
\begin{gather}
\begin{split} 
   \mu_i(t)&=\sum_{j=1}^{l_i}
M_1M_{02}M_{03}M_{04}M_5M_{06}M_7M_8....M_K
             = \sum_{j=1}^{l_i}\prod_{k=0}^{K} M_{0k}^{1-h{i,k}}M_k^{h_{i,k}}\\
h_{i} & = (h_{i,1},h_{i,2},...) = ( 0,0,0,1,0,0,1,0...) 
 \hspace{1cm} \sum_{k=1}^{K}h_{i,k} =Kz
\end{split}
\end{gather}
Here, $K$ is the number of the layers of time scale and these layers respectively include each source of information that is available for traders.
$l_i$ is the number of lines including only logical products.
 For simplicity, we set $l_i=1$. 
A trader uses the information of $Kz$ sources, because he uses only $z$ number of sources whose corresponding element of his $h$ is equal to 1.
$M_{0k}$ is a fixed stochastic variable and is always takes $m_0=0.5$.
$M_k$ takes either $m_1=0.9$ or $m_2=0.1$ in random order, at the intervals of $\frac{T}{2^K}$\\
\textbf{Market Price}\\
We suppose that traders make an order by the equation (3) and all orders are traded. Then the trading frequency and volume of the market is in proportion to following equation.
\begin{gather}
\begin{split} 
   \mu(t)&=\frac{1}{N}\sum_{i=1}^{N}\mu_i(t)\\
\end{split}
\end{gather}
Here, $N$ is the number of traders.
Like the MMAR model, every unit of the trading time is set every time a trade is completed. In this case the trading time is proportional to trading frequency.
The price is assumed to behave in fractional Brownian motion (FBM) along this trading time. Here, $H=0.5$ for which FBM is normal Brownian motion.\\
\textbf{Extraction of the Macro Model}\\
Using the trading frequency in an time interval (volume)
$\mu(t)$, Hoolder exponents, scaling functions and multifractal spectrums are calculated as follows. 
\begin{gather}
\begin{split} 
\mu(t)&=\frac{1}{N}\sum_{i=1}^{N}\prod_{k=0}^{K}M_{0k}^{1-h_{i,k}}M_k^{h_{i,k}}=\frac{1}{N}\sum_{i=1}^{N}\prod_{k=0}^{K}\sum_{h_k=0}^{1}\delta(h_{i,k}-h_k)M_{0k}^{1-h_k}M_k^{h_k}\\
&=\sum_{h_1=0}^{1}\sum_{h_2=0}^{1}\sum_{h_3=0}^{1}
\rho(h_1,h_2,h_3,...)\prod_{k=0}^{K}M_{0k}^{1-h_k}M_k^{h_k}\\
\rho(h_1 & ,h_2,h_3,...)=\frac{1}{N}\sum_{i=1}^{N} \prod_{k=0}^{K}\delta(h_{i,k}-h_k)
\underset{N\rightarrow\infty}{\sim}\prod_{k=0}^{K}\frac{1}{N}\sum_{i=1}^{N}\delta(h_{i,k}-h_k)
=\prod_{k=0}^{K}\rho(h_k)\\
\rho(1)&= \frac{1}{N}\sum_{i=1}^{N}\delta(h_{i,k}-1)=z  \qquad 
\rho(0)= \frac{1}{N}\sum_{i=1}^{N}\delta(h_{i,k})=1-z\\
\mu(t)
&=\prod_{k=0}^{K}\sum_{h_k=0}\rho(h_k) M_{0k}^{1-h_k}M_k^{h_k}
=\prod_{k=0}^{K}(M_{0k}(1-z)+M_kz)\\
\mu_n&=(m_0(1-z)+m_1z)^n(m_0(1-z)+m_2z)^{(K-n)}  \quad 
N_n={}_KC_n\\
\end{split}
\end{gather}
\textbf{Hoolder exponents}
\begin{gather}
\begin{split} 
\alpha &=\frac{\log(\mu_n)}{\log{2^{-K}}}  =-\frac{n}{K}\alpha_1
-(1-\frac{n}{K})\alpha_2\\
\alpha_1 & =\log_2(m_0(1-z)+m_1z) \qquad 
\alpha_2  =\log_2(m_0(1-z)+m_2z)
\end{split}
\end{gather}
\textbf{scaling functions}($N>>1$,$K>>1$)
\begin{gather}
\begin{split} 
\tau_\theta(q) & =-\log_2[(m_0(1-z)+m_1z)^q+(m_0(1-z)+m_2z)^q]
\quad \textrm{(frequency)}\\ 
\tau(q) &=H(\tau_{\theta}(q)+1)-1 \hspace{5.4cm}\textrm{(return)}
\end{split}
\end{gather}
\textbf{multifractal spectrum}($N>>1$,$K>>1$)
\begin{gather}
\begin{split} 
f_\theta(\alpha)&=\frac{\log(N_n)}{\log(2^K)} 
=-\frac{\alpha-\alpha_2}{\alpha_1-\alpha_2}\log_2(\frac{\alpha-\alpha_2}{\alpha_1-\alpha_2})
-\frac{\alpha_1-\alpha}{\alpha_1-\alpha_2}\log_2(\frac{\alpha_1-\alpha}{\alpha_1-\alpha_2})\\
& \hspace{7.8cm} \textrm{(frequency)}\\
f(\alpha)&=f_\theta(\alpha/H) \hspace{6cm}\textrm{(return)}
\end{split}
\end{gather}
\vspace{0cm}
\section{Simulation}
\begin{figure*}[hb]
\begin{center}
\includegraphics[scale=0.4]{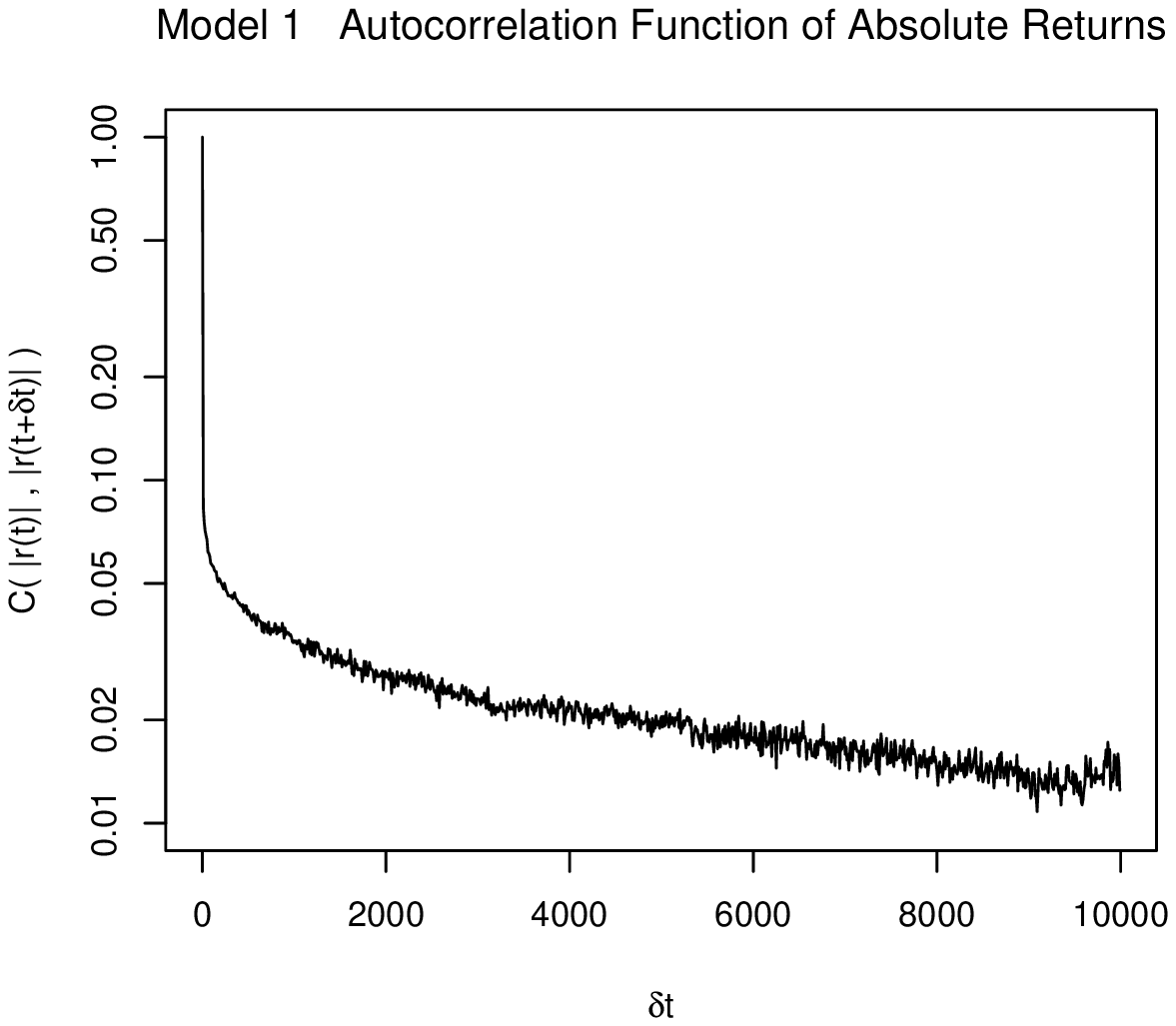}
\includegraphics[scale=0.4]{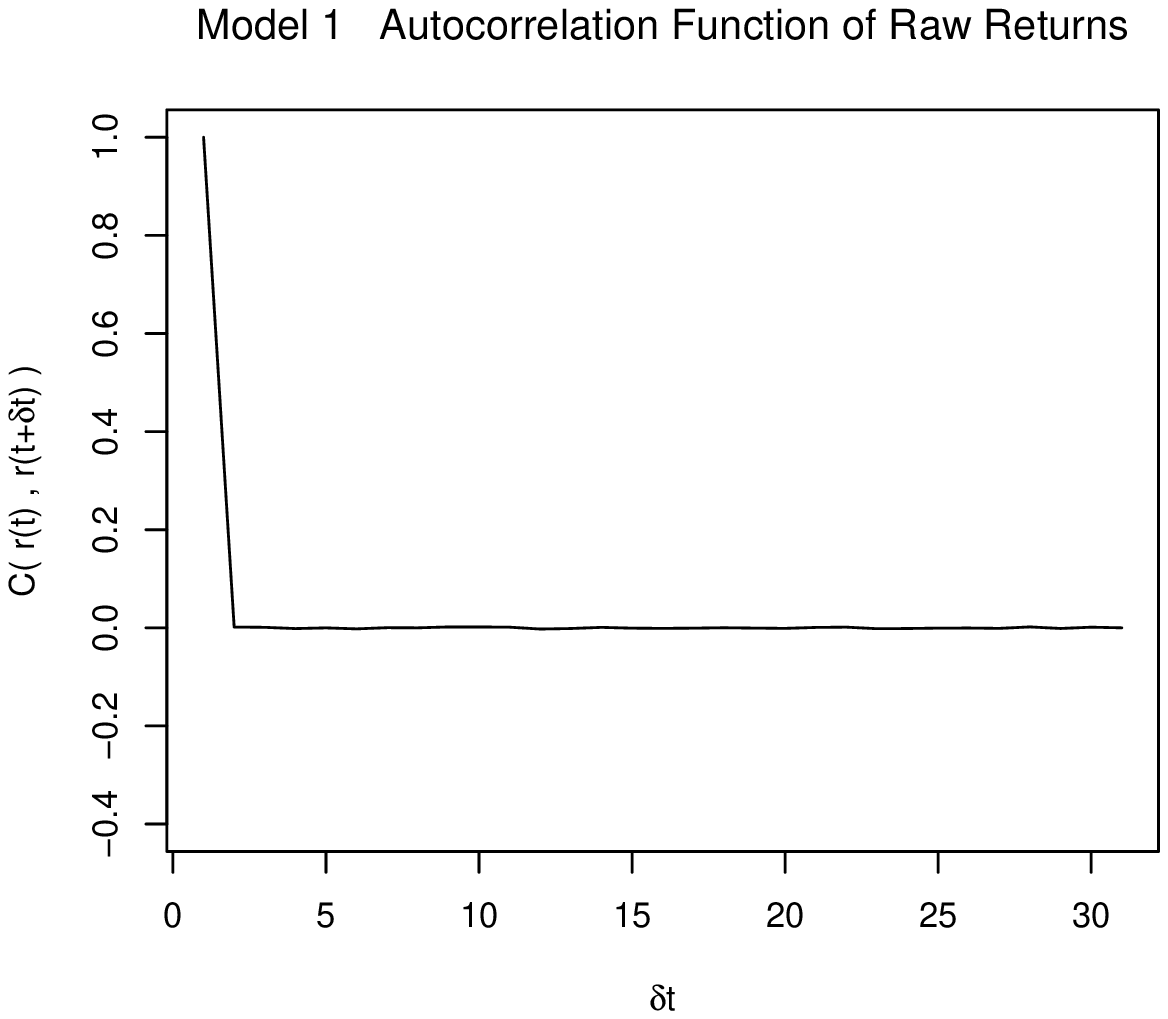}\\
\caption{stylized facts(market simulation)}
\label{stylizedfact}
\includegraphics[scale=0.4]{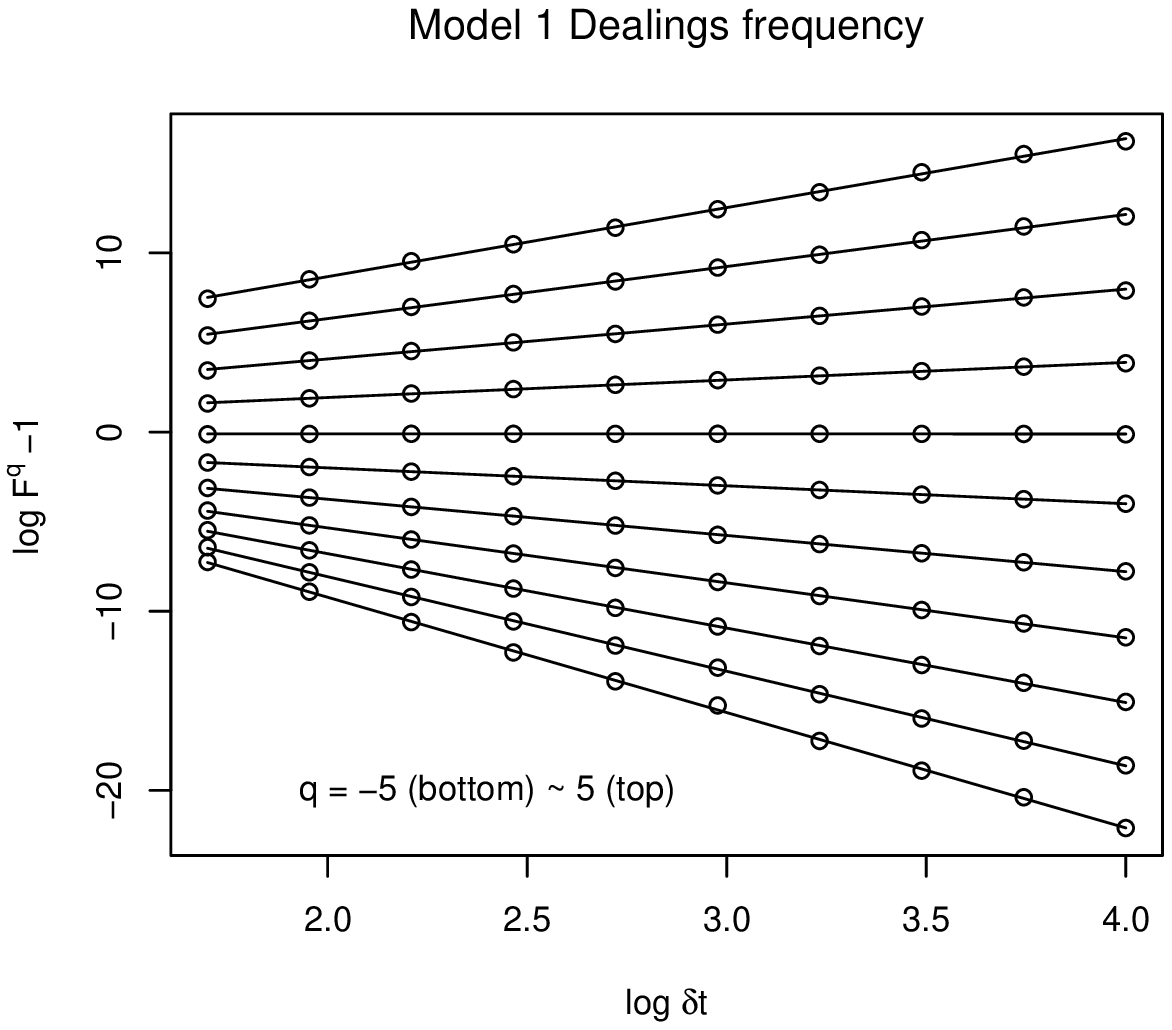}
\includegraphics[scale=0.4]{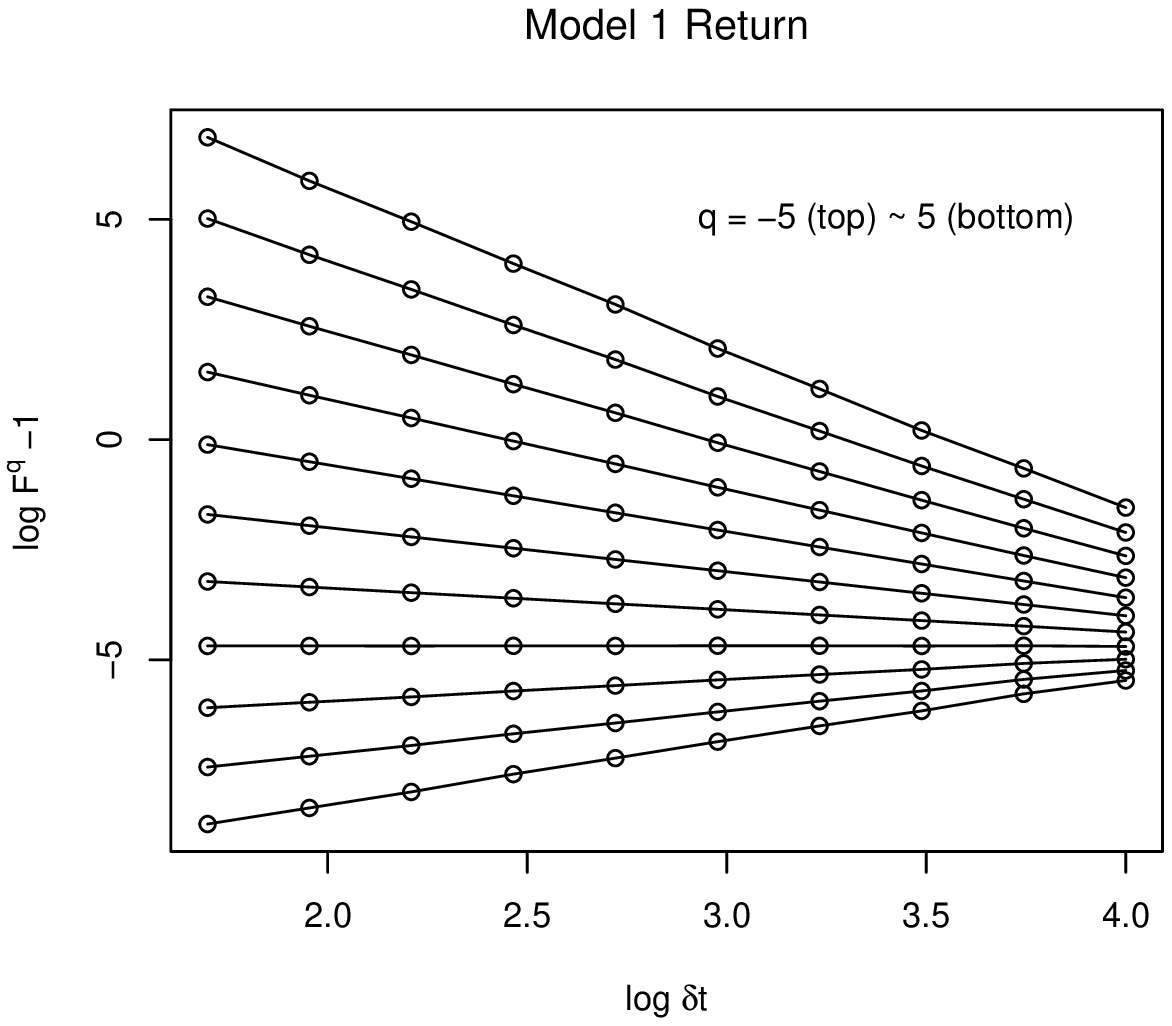}\\
\includegraphics[scale=0.4]{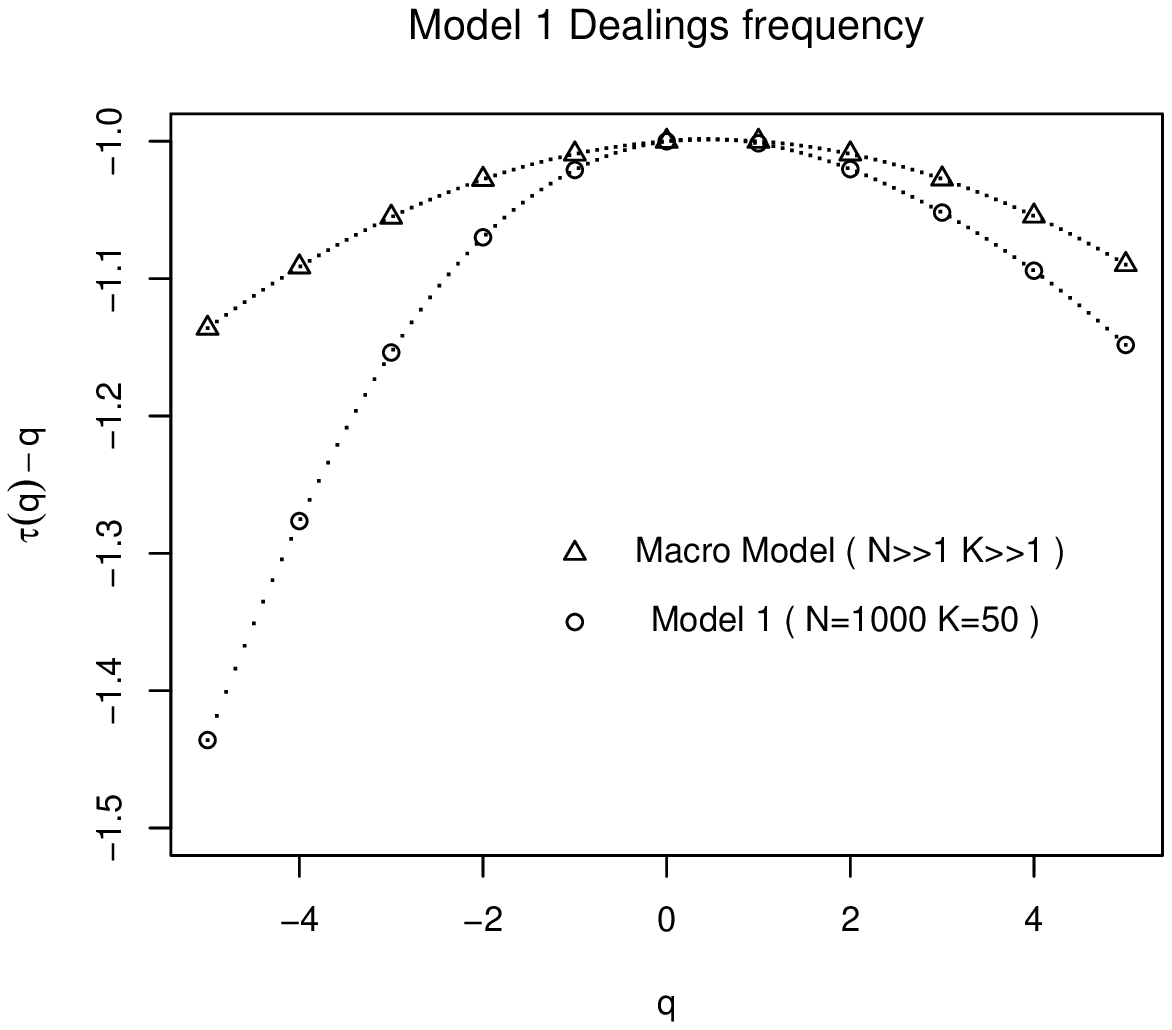}
\includegraphics[scale=0.4]{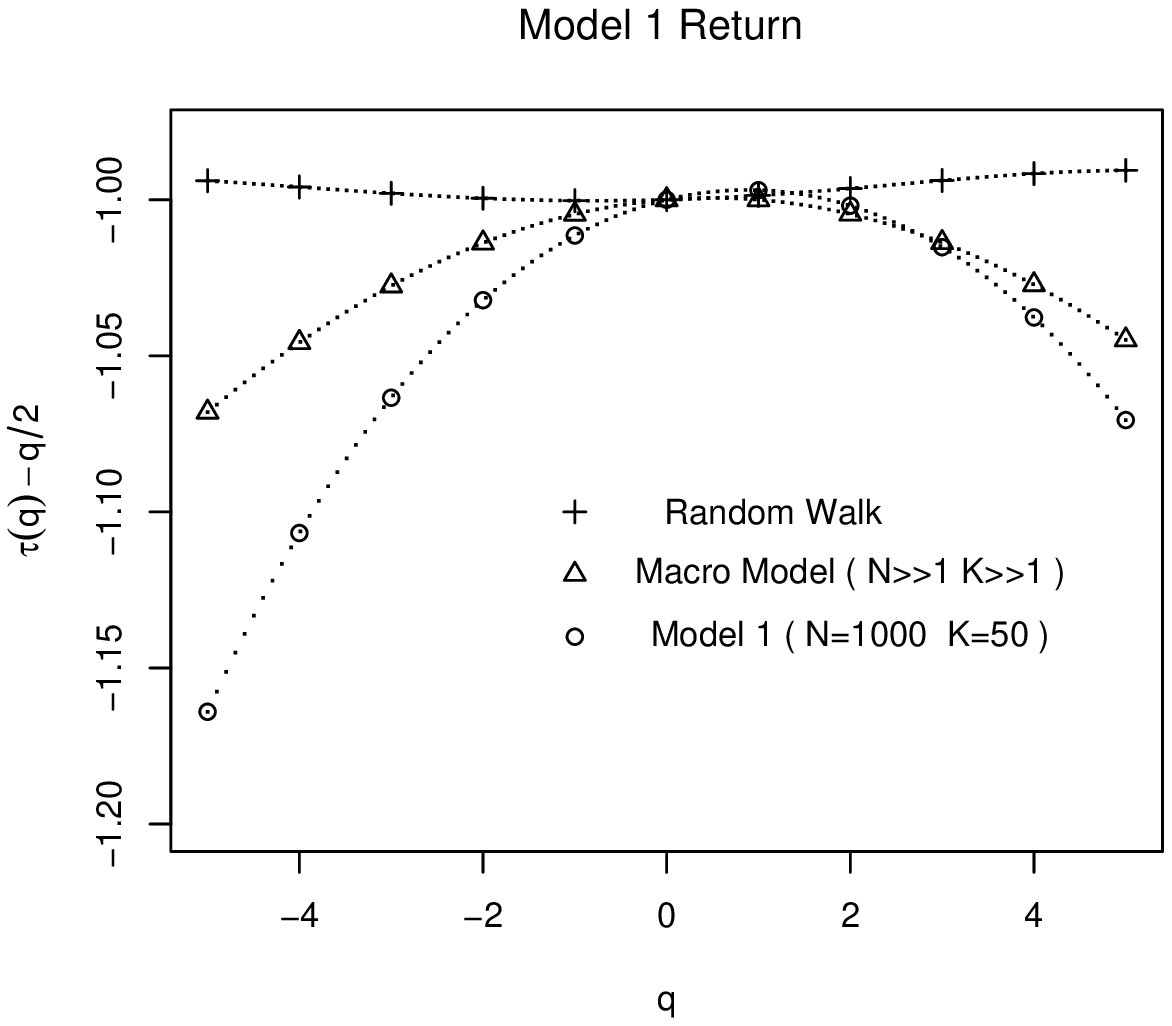}\\
\includegraphics[scale=0.4]{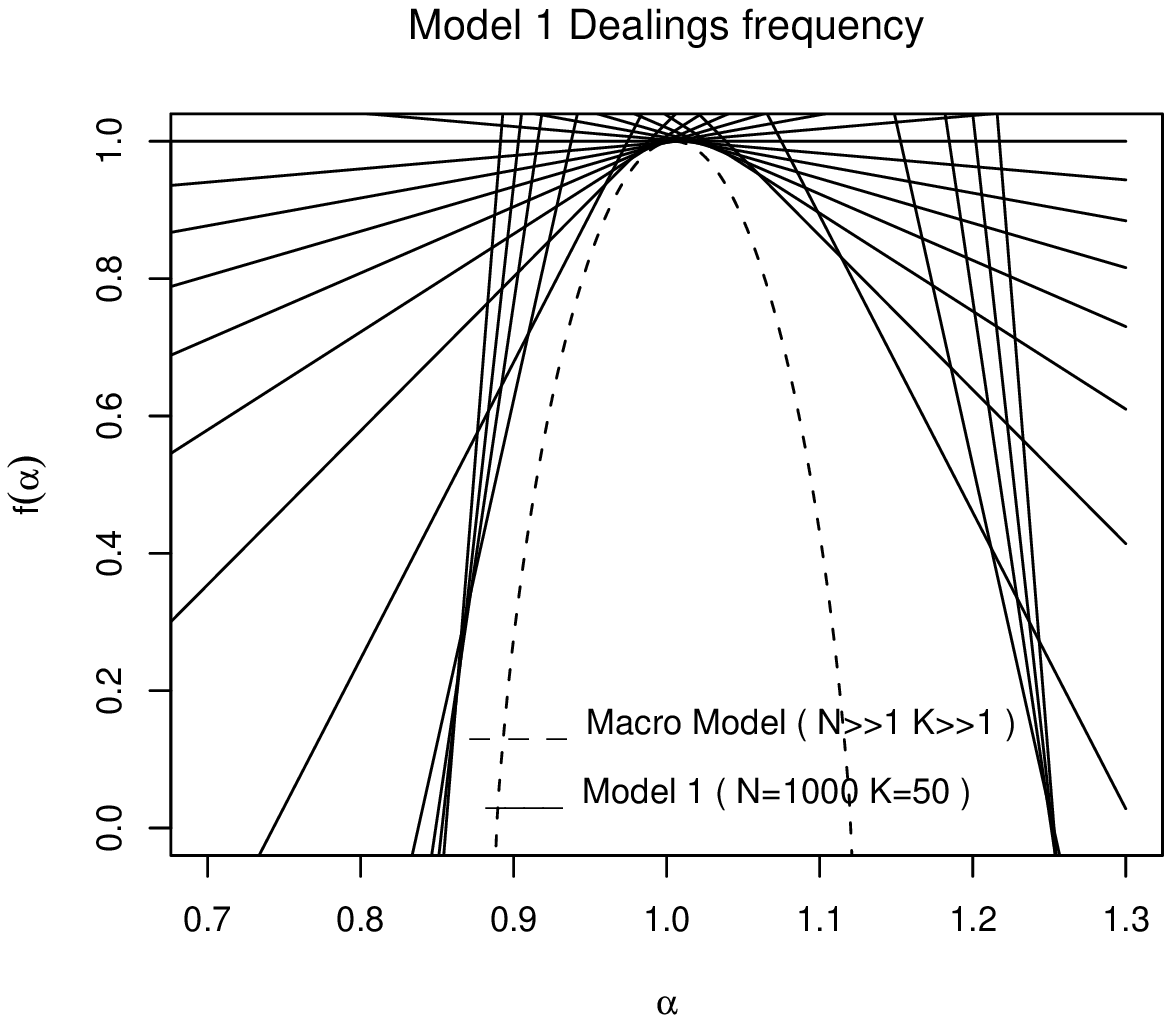}
\includegraphics[scale=0.4]{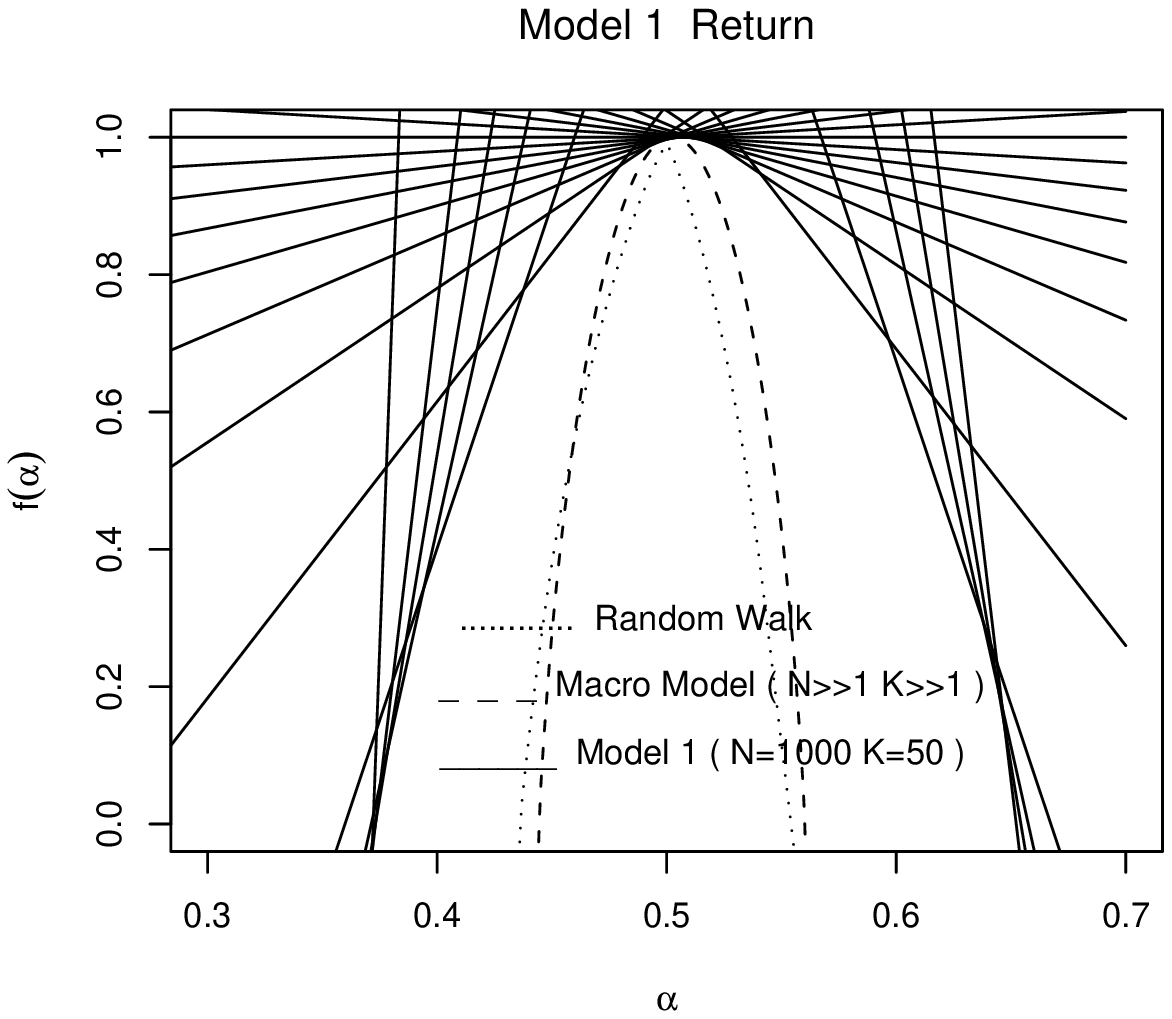}\\
\caption{Multifractal Property}
The comparison of market simulation, macro model and random walk.
\label{multifractal}
\end{center}
\end{figure*}

We show the results of the micro model and the macro model.
There are some studies that state normal moment estimation lacks accuracy, so we adopted the MF\_DFA (MultiFractal Detrended Fluctuation Analysis) method \cite{Kantelhardt1} for multifractal analysis.
The parameters used were
$N=1000,K=50,T=10000,m_1=0.9,m_2=0.1,z=0.1$. We generated time series in the length of $10^7$.\\
Fig. \ref{stylizedfact} shows the stylized facts of speculative markets.
Fig. \ref{multifractal} (top) show $q$th moments of frequency (left) and return (right) in $log-log$ scale.
They are located along the straight lines, i.e. they show fractal properties.
Fig. \ref{multifractal} (middle) show scaling functions of frequency (left) and return (right).
They show the multifractal property stronger in the market simulation than in the macro model.
Fig.\ref{multifractal} (bottom) show multifractal spectrums. 
In this case, it appears the difference between random walk and the macro model is small, because in the left and right side of the graph, large values of $|q|$ dominate.
The result of the market simulation clearly shows multifractal property.
%
\section{Conclusion}
We Proposed an example of a market simulation (micro model) that showed multifractal properties.
The characteristic aspect of our model is that each trader makes his decision by the logical products of a few probabilities, which represent each time scale respectively and shows the appropriateness or accuracy of a source of information.
From this market simulation model, we can extract the MMAR model as a macroscopic stochastic equation for the limit of $N>>1,K>>1$.
This simple simulation can make clear what is the cause of each stylized fact, and which facts come from the same cause.
\begin{enumerate}
\item Fat-tail depends on the kind of multifractal cascade of trading time.
\item Long-memory is caused by multifractal cascades.
\item Short-memory is caused by the Brownian motion of price along the trading time.
\item Multifractal property is caused by multifractal cascades.
\item The layers of a time scale originates from the time scale layers used in our daily lives.
\item The multifractal cascade originates from each trader's strategy through sources of information representing each time scale.
\end{enumerate}
\bibliographystyle{plain}
\bibliography{market.bib}
\end{document}